# DocEDA: Automated Extraction and Design of Analog Circuits from Documents with Large Language Model

Hong Cai Chen
School of Automation
*Southeast Universisity*
Nanjing, China
chenhc@seu.edu.cn

Longchang Wu
School of Automation
*Southeast Universisity*
Nanjing, China
1149670719@qq.com

Ming Gao
School of Automation
*Southeast Universisity*
Nanjing, China
220221874@seu.edu.cn

Lingrui Shen
School of Automation
*Southeast Universisity*
Nanjing, China
220242150@seu.edu.cn

Jiarui Zhong
School of Automation
*Southeast Universisity*
Nanjing, China
220245143@seu.edu.cn

Yipin Xu
School of Automation
*Southeast Universisity*
Nanjing, China
1455120321@qq.com

*Abstract*—Efficient and accurate extraction of electrical parameters from circuit datasheets and design documents is critical for accelerating circuit design in Electronic Design Automation (EDA). Traditional workflows often rely on engineers manually searching and extracting these parameters, which is time-consuming, and prone to human error. To address these challenges, we introduce DocEDA, an automated system that leverages advanced computer vision techniques and Large Language Models (LLMs) to extract electrical parameters seamlessly from documents. The layout analysis model specifically designed for datasheet is proposed to classify documents into circuit-related parts. Utilizing the inherent Chain-of-Thought reasoning capabilities of LLMs, DocEDA automates the extraction of electronic component parameters from documents. For circuit diagrams parsing, an improved GAM-YOLO model is hybrid with topology identification to transform diagrams into circuit netlists. Then, a space mapping enhanced optimization framework is evoked for optimization the layout in the document. Experimental evaluations demonstrate that DocEDA significantly enhances the efficiency of processing circuit design documents and the accuracy of electrical parameter extraction. It exhibits adaptability to various circuit design scenarios and document formats, offering a novel solution for EDA with the potential to transform traditional methodologies.

*Keywords*—Electronic Design Automation (EDA), Large Language Models (LLMs), Document Parsing, Chain-of-Thought (CoT), Netlist Generation.

I. INTRODUCTION

The rapid advancement of Electronic Design Automation (EDA) has drastically transformed how engineers design and simulate complex electronic circuits [1,2]. A crucial challenge within this domain is the accurate and efficient extraction of electrical parameters from circuit datasheets and design documents [3]. Historically, this process has been performed manually, which is both time-consuming and prone to human error. As circuit designs become increasingly sophisticated, the need for automated solutions that enhance both the speed and accuracy of data extraction has grown significantly.

To address these challenges, we introduce DocEDA, a novel automated system that leverages cutting-edge artificial intelligence techniques to streamline the extraction of critical electrical parameters from a wide array of circuit documentation and optimizing the circuit. DocEDA integrates the capabilities of advanced computer vision methods and Large Language Models (LLMs) to significantly enhance the efficiency of traditional workflows. The system contains five key parts:

**1) Document Layout Analysis**: We propose a specialized layout analysis model that effectively dissects circuit datasheets and technical documents into distinct sections, including parameters, circuit diagrams, performance curves, etc. This enables the precise extraction of relevant information, forming the foundation for subsequent tasks.

**2) Parameter Extraction Using CoT**: By harnessing the Chain-of-Thought reasoning capabilities of LLMs, DocEDA automates the extraction of critical component parameters from both textual and tabular formats. Our approach integrates document retrieval, iterative refinement, and dynamic preference optimization, providing accurate and structured data for circuit simulation.

**3) Database**: The database stores structured electronic component and circuit design data derived from document layout parsing and parameter extraction. The designed data structure enables rapid query and design optimization, while the structured content within this database provides substantial data for training other models.

**4) Circuit Diagram Analysis**: This model combines component recognition with topology recognition enabling the system to accurately extract circuit topologies and generate standard-compliant netlists, a vital component for seamless integration with circuit simulation tools.

**5) Circuit Layout Optimization**: The system incorporates a dynamic optimization framework that refines circuit layouts, beginning with an initial schematic and evolving towards an optimized layout design. By applying space mapping, EM simulations are replaced with schematic simulations which significantly reduce the optimization time.

DocEDA adapts to various circuit design scenarios and document formats, offering an innovative solution for electronic design automation with the potential to transform traditional methodologies. In the following sections, we will provide a detailed overview of these five key models.

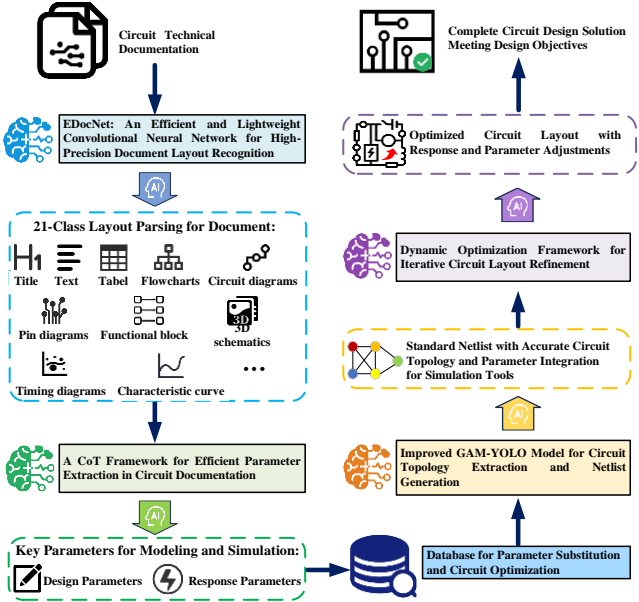

Fig. 1 The whole framework of DocEDA framework.

## II. THE PROPOSED DOCEDA FRAMEWORK

The architecture of our DocEDA is shown in Fig. 1. Initially, it employs a specialized document layout analysis model to classify and segment input circuit datasheets and technical documents into sections like model numbers, tables, parameters, circuit diagrams, and performance curves, laying the groundwork for parameter extraction and circuit design. DocEDA then leverages the Chain-of-Thought reasoning capabilities of Large Language Models (LLMs) to automatically extract key component parameters from these sections. Accurately identifying parameters in both text and tables, the system generates structured data stored in a comprehensive database of electronic components and circuit information. This database supports subsequent circuit design and simulation and aids in training large models to enhance system performance. Next, using an improved GAM-YOLO model combined with topology recognition techniques, it precisely extracts the circuit's topological structure. By integrating the circuit diagrams with the extracted parameters, it generates standard-compliant netlists for circuit simulation tools. Finally, a dynamic optimization framework is introduced to iteratively refine the circuit design. Utilizing the initial schematic, generated netlists, and parameter data, the system enhances the layout through a Bayesian optimization and space-mapping method to obtain an optimized layout design efficiently. This workflow significantly improves the efficiency of processing circuit design documents, reducing reliance on manual intervention.

## III. METHODOLOGY

### A. Document Layout Analysis

The existing document parsing models, such as LayoutLM [4], Dit [5], etc., has limited capacity to analyze the specific characteristics of the document related to electronic components and circuit design. Parsing electronic component documentation is complex due to the vast number of components, and their varying attributes like shape, size, color, and markings. The documents are packed with model numbers, parameters, specs, and functions, often presented in various formats including text, symbols, and tables. To address this limitation, this paper proposes the EDocNet model [6] specifically designed for electronic datasheets and design documents.

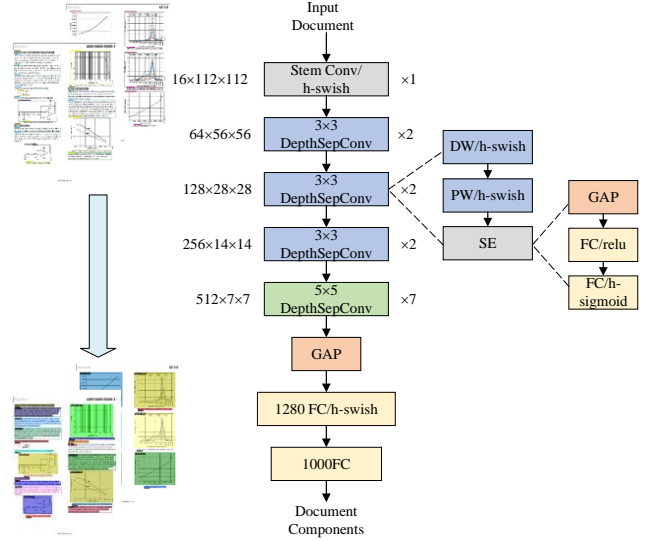

Fig. 2 The basic structure of document paring and the EDocNet network.

The EDocNet model is based on a multi-layer depthwise convolutional neural network. It is trained using both focal and global distillation techniques to enhance its ability to capture fine-grained details and global patterns, thereby improving the accuracy of document feature extraction. To meet the specific parsing needs of electronic device documentation, we developed a specialized dataset containing 21 categories, enabling the model to precisely recognize and classify various types of information within these documents. As the network structure shown in Fig. 2, EDocNet employs deep separation convolution to reduce convolution computation and introduces the SE module (Squeeze and Excitation) to enhance feature representation capability. Additionally, lightweight activation function H-swish and optimized network structure are used, allowing EDocNet to achieve a good balance between speed and accuracy.

In the model training, the FGD method is used, which is a knowledge distillation method combining local and global information to improve the feature learning ability of neural networks. Focus distillation enhances understanding of local details by directing student models to focus on key regions or fine-grained features in the input data; Global distillation, on the other hand, helps models capture broader context information and ensures global consistency. This dual strategy not only guarantees the local accuracy, but also enhances the overall feature representation ability of the model, which is especially suitable for tasks that need to pay attention to details and global structure at the same time.

### B. Parameter Extraction Using CoT

During the Layout Analysis phase of circuit technical documentation, the document is segmented into several circuit related contents. These modules are represented in three main formats: text, tables, and images. The subsequent task is to extract relevant technical parameters and key data from these information types to support further analysis. To achieve this, we apply a Chain of Thought approach [7], utilizing large language models (LLMs) to guide the efficient extraction of parameters from the document. This section introduces two key methods—Iterative Refinement Optimization (IRO) and Priority Ordering (PO). These methods work in tandem to enhance the search and extraction process, improving both accuracy and efficiency. Specifically, IRO iteratively refines the search strategy to pinpoint relevant parameters with increasing precision, while PO prioritizes text blocks, enabling the model to focus on the most pertinent information. Together, these methods provide an effective solution for extracting technical parameters from complex document structures. The following sections will detail their implementation and advantages.

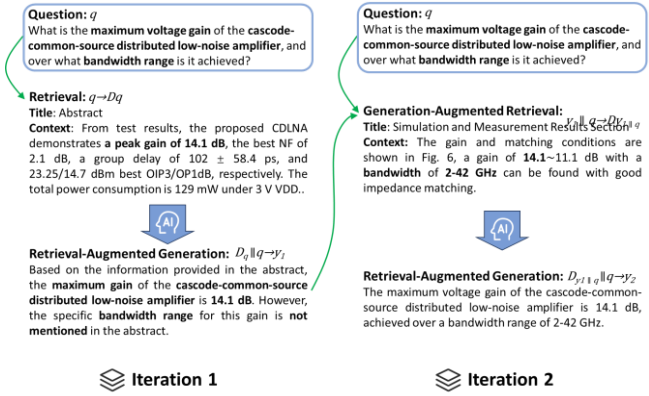

Fig. 3. Iterative Parameter Retrieval and Inference Process with IRO: A Concrete Example.

**Iterative Retrieval Optimization (IRO)**: Traditional retrieval methods often fall short due to the semantic gap between the user's query and the required knowledge [8]. IRO addresses this challenge by iteratively alternating between retrieval and generation processes, progressively narrowing this gap, and improving parameter extraction accuracy. In each iteration, the model leverages the output from the previous generation step, combining it with the original query to refine the retrieval process. Specifically, given a user query $q$ and a corpus $D = \{d\}$, the IRO method performs the following steps in the $t$-th iteration:

**(1) Retrieval Stage:** Concatenate the previous output $y_{t-1}$ with the query $q$ to form a new query. Use this to retrieve relevant documents from the corpus $D$, resulting in a retrieval set $D_{y_{t-1}\|q}$.

**(2) Generation Stage:** Utilize an LLM $M$ to generate a new output $y_t$ based on the retrieved documents $D_{y_{t-1}\|q}$ and the original query $q$.

This iterative process is formalized as:
$$y_t = M(y_t | prompt(D_{y_{t-1}\|q}, q)), \quad 1 \leq t \leq T \quad (1)$$

where $T$ is the maximum number of iterations, and $y_T$ is the final output provided by the model [9]. Fig. 3 depicts the IRO workflow, illustrating how each iteration refines the retrieval and generation processes.

The Priority Ordering (PO) method aims to enhance the parameter extraction within circuit research papers by defining the priority of knowledge blocks obtained from the preliminary layout recognition. Through structured analysis, the model is able to identify and label different knowledge blocks, such as "Abstract," "Introduction," "Circuit Design," "Measurement Results," and "Simulation Results." These knowledge blocks form a hierarchical structure that provides a clear path for subsequent searches. Leveraging expert knowledge of typical parameter locations, the system assigns varying priority levels to each knowledge block. For example, critical modeling parameters are often concisely recorded in the "Abstract", thus this section is assigned the highest priority. By applying constraint optimization theory, the search process is primarily conducted within high-priority sections, and consider lower-priority sections only when necessary. Fig. 4. shows preference optimization framework for circuit parameter extraction in knowledge blocks.

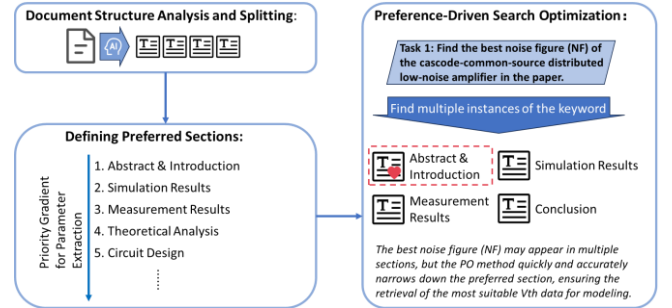

Fig. 4. Preference Optimization Framework for Circuit Parameter Extraction in Knowledge Blocks.

*C. Database*

The database is designed to comprehensively store and manage structured data parsed from electronic design documents, meeting the needs of electrical parameter extraction, document layout analysis, and large model training. The database comprises multiple tables to store different types of critical data.

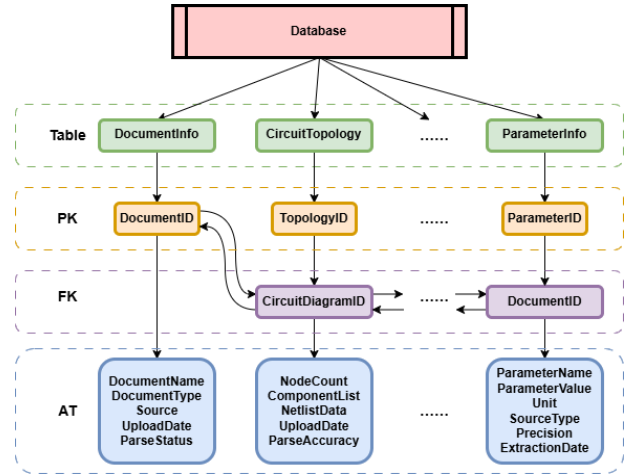

Fig. 5. The data structure of the database.

The Document Information Table records the essential metadata for each circuit document, including document ID, source, and parsing status, etc., enabling the management and tracking of all documents. For extracted parameters, the database includes a Parameter Table, which stores each parameter's name, value, unit, and other details, ensuring that these data can be efficiently referenced in subsequent analyses. The Layout Information Table accurately categorizes document layouts, recording the category (e.g., model number, table, circuit diagram, performance curve) and precise location within the document (page number and coordinates). The Circuit Topology Table is dedicated to storing the parsed results of circuit diagrams, including node, component list, connection, and generated netlist, enabling seamless integration with simulation tools. The Image Data Table stores all image data within documents, including image type, annotated region coordinates, captions, etc., making it suitable for further analysis. Finally, the Label Table records label content associated with various layout elements, assisting the model in associating labels with document content.

The relationships among these tables ensure that document information, parameter data, layout information, images, and labels are efficiently managed within the system, facilitating quick queries, parameter substitutions, circuit design optimization, and providing rich structured data support for model training. This enhances the overall efficiency of data management and analysis.

*D. Image-to-Circuit Process*

The circuit diagram obtained from documents are in the image form. To use them in circuit design, it is preferred transform them into circuit netlists. This transformation process consists of two main steps: circuit component recognition and circuit topology recognition. The overall method is described in Fig. 6.

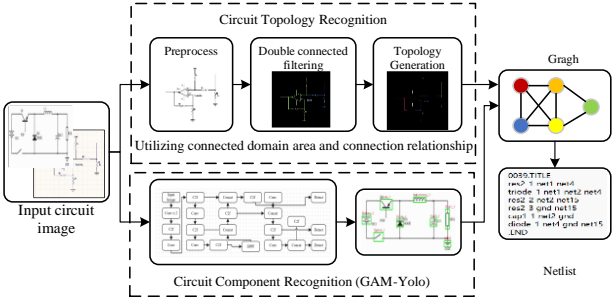

Fig. 6. Overall method of image-to-circuit process.

**(1) Circuit component recognition**

This paper proposed GAM-YOLO [10], an enhanced Attentional YOLOv8 model tailored for circuit component recognition. YOLOv8 [11], a state-of-the-art (SOTA) object detection model, is composed of three main architectural components: the backbone, the neck, and the head. To improve the detection of components within circuit diagrams, which display unique foreground and spatial features, GAM-YOLO incorporates the Global Attention Module (GAM) [12] between the neck and the head of the YOLOv8 network. The GAM mechanism effectively combines channel and spatial attention, allows the network to weight or adjust the feature maps before generating bounding boxes and class predictions, thereby enhancing the feature representation of the regions of interest. thereby enhancing the model's ability to recognize circuit components. By doing so, the method can enhance the network's ability to detect circuit components under complex scenes and small targets.

**(2) Circuit topology recognition**

Circuit topology recognition involves two main stages: gray-scaling and binarization. For light-background circuit diagrams, Otsu's method is preferred for binarization as it effectively removes scattered points and grid-like backgrounds without introducing noise, which is more efficient than the triangle method. In contrast, for dark-background diagrams, the triangle method is more appropriate for binarization as it preserves dark wire details, with any noise being manageable through post-processing.

After preprocessing, a two-step connected domain filtering algorithm is applied to extract net information vital for determining circuit topology. The first step filters out small connectivity domains, typically less than 10% of the image's total pixel count, removing text and noise while retaining the main circuit components. The second step refines this by integrating component location data, discarding any connectivity domains that do not intersect with the predicted bounding boxes of components. The remaining connectivity domains represent the circuit's nets.

After the extraction of both component and net information from the circuit diagram, intersection detection can establish the topological relationships within the circuit. This method provides a detailed understanding of the circuit's structure, which is crucial for subsequent analysis and design processes.

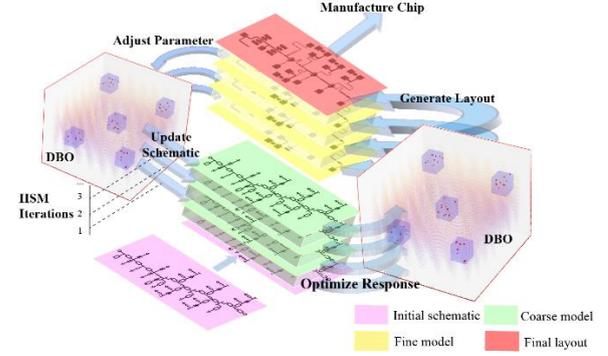

Fig. 7. Block diagram of proposed circuit optimization method.

*E. Circuit Layout Optimization*

After obtaining circuit netlist and parameters, the circuit can be automatically simulated and optimized. As layout optimization requires time-consuming EM simulations, the optimization should be designed with least EM simulations. The circuit diagram optimization framework, as depicted in Fig. 7, accepts the initial circuit schematic as input and automatically yields an optimized layout suitable for chip fabrication. This approach employs Implicit Space Mapping (ISM) for iterative optimization, where each iteration involves a coarse model representing the schematic and a fine model representing the layout [13]. Both the coarse and fine models incorporate the same design parameters $x_c$, with the coarse model also including additional auxiliary parameters $x_a$. The initial schematic serves as the coarse model for the first iteration. The process of the $i$-th ISM iteration are as follows:

All auxiliary parameters $x_a(i)$ are fixed and the coarse model is optimized for optimal response to obtain the parameters $x_c^*(i)$ and the response $F_c(i)$, based on Eq. (1). Then, based on $x_c^*(i)$, generate the circuit layout and perform layout simulation to obtain the fine model response $F_f(i)$.

$$x_c^*(i) = \arg\min_{x_c} F_c(x_c, x_a^*(i)) \quad (1)$$

Then, with all design parameters $x_c^*(i)$ fixed, the auxiliary parameters $x_a$ in the schematic are adjusted so that response deviation of two models converges to zero based on Eq. (2), resulting in refreshment of $x_a(i+1)$ in coarse model.

$$x_a(i+1) = \arg\min_{x_a} |F_c(x_c^*(i), x_a) - F_f(x_f(i))| \quad (2)$$

After a few iterations, the responses of the coarse model and the fine model are consistent. Since the coarse model has been optimized to the target response and the fine model has also achieved the optimal response, $x_c$ in the last iteration is used as the design parameter to achieve the optimal response of the layout.

The process of response optimization and parameter adjustment, which involves minimizing Eq. (1) and Eq. (2), is a high-dimensional black-box optimization problem. Dynamic Bayesian Optimization (DBO) is used for Efficiently obtaining the global optimum of a function. In DBO, A multi-trust region-based batch BO (mTURBO) algorithm is used for global search [14]. Global search parallelizes the training of Gaussian Process (GP) in multiple regions, using Thompson Sampling (TS) to obtain candidate points. The boundary of the regions is dynamically adjusted based on sample value to explore the global optimum. When global regions converge, Expected Coordinate Improvement BO (ECIBO) is used as a local search to accelerate local convergence speed and adjust global search area [15]. ECIBO measures the potential improvement along a particular coordinate, shifting high-dimensional problems to low-dimensional ones, showing high efficiency when searching within a small range in high dimensions.

IV. PARAMETER EXTRACTION RESULTS ANALYSIS

To validate the effectiveness of the proposed framework, the performance of each model is presented with testing results for the dataset and an example of low noise amplifier design paper.

*A. Dateset*

The dataset for this paper is derived from 8,766 electronic component documents or datasheets and has been manually annotated to categorize the content into 21 categories, including Text, Title, List, Table, functional block diagrams, flowcharts, characteristic curve diagrams, timing diagrams, circuit diagrams, pin diagrams, engineering drawings, sampling diagrams, 3D schematics, pin name diagrams, marking diagrams, appearance diagrams, functional register diagrams, layout diagrams, data structure diagrams, and other parts diagrams. Meanwhile, a paper [16] on low noise amplifier (LNA) is parsed as the example of the whole framework, whose design is optimized automatically.

*B. Document Layout Analysis Result*

The document layout analysis was carried out on the dataset. For comparison, state-of-art methods including Yolov8, LayoutLMv3 and Dit are also evaluated. The performance indexes of the evaluation are Average Recall (AR) and Average Precision (AP), and the results are shown in Table I.

TABLE I MODEL COMPARISON RESULT

| Model name | AP | AR | Time |
|---|---|---|---|
| Yolov8 | 0.684 | 0.675 | 1.9s |
| LayoutLMv3 | 0.563 | 0.944 | 0.3s |
| Dit | 0.507 | 0.908 | 0.6s |
| EDocNet | 0.765 | 0.945 | 1.2s |

The results in the table demonstrate that the model excels in both Average Precision (AP) and Average Recall (AR) metrics. Notably, there is a substantial increase in the recall rate, with AR reaching 0.945, which signifies that the model is adept at identifying targets even when there is a high volume of detections. This capability is precisely what is required for the model to be effectively utilized in the context of electronic device documentation. The LNA paper is parsed using EdocNet and the results are shown in Fig. 8.

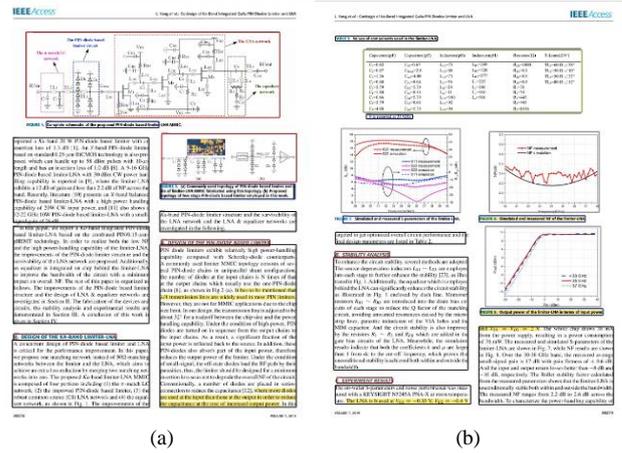

Fig. 8. Parsing results for the LNA paper. All contents are detected.

Fig. 9. The results of parameter extraction using CoT.

*C. Chain-of-Thought (CoT) Result*

To validate the effectiveness of the CoT approach in extracting parameters from circuit technology documents, we selected a paper focusing on a Ka-band integrated PIN diode limiter and low-noise amplifier (Limiter-LNA). Using our proposed CoT-based parameter search strategy, integrated within the DocEDA framework, we conducted the analysis. The interaction process between the user and the agent is illustrated in Fig. 9. By analyzing the paper, the system first identifies the circuit type and its functionality. Subsequently, it precisely searches for the design parameters and experimentally measured response parameters within the paper and presents the results in a structured format, enabling further analysis and informed decision-making.

## D. Image-to-Circuit Result

To evaluate the effectiveness of the image-to-circuit conversion process, a testing dataset consisting of 200 circuit diagrams sourced from real-world industrial environments was assembled and tested. The generated circuit graphs were manually inspected for precision to ensure the accuracy of the conversion. The results of this evaluation are summarized in Table II. The transformation accuracy is larger than 90% ensuring it can be applied in practical engineering. Additionally, an illustrative example of the process is provided in Fig. 10, which demonstrates the conversion of circuit diagram in the studied paper [16].

TABLE II  MODEL PERFORMANCE RESULT

| Number of tested circuit diagrams | 200 |
|---|---|
| Image-to-circuit accuracy | 90.5% |

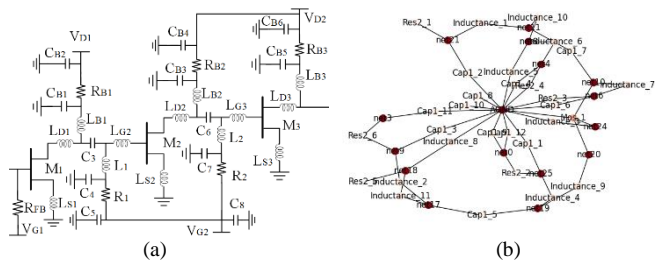

(a)   (b)

Fig. 10. The results of image-to-circuit transform. (a) Circuit image. (b) Transformed circuit gragh representation.

TABLE III  COMPARISON WITH OTHER LNA WITH PIN DIODE LIMITER

| Item | Process | Freq (GHz) | OP1dB (dBm) | $P_{dc}$ (mW) | Gain (dB) | Noise figure | Area (mm$^2$) |
|---|---|---|---|---|---|---|---|
| 1 | 90-nm GaN | 30-38 | NA | 76 | 17 | <2.6 | 2.5×1.5 |
| 2 | MSAG | 30-36 | >11 | 72 | 18.5 | <2.5 | 2.5×1.2 |
| 3 | GAN | 15-34 | NA | 300 | 18.5 | 0.8-1.5 | 2.0×1.3 |
| 4 | 0.1-μm GaAs | 28-38 | NA | NA | 21 | 2.5 | NA |
| 5 | 0.1-μm GaAs | 30-38 | NA | NA | 18 | 2.3 | NA |
| This work | 0.1-μm GaAs | **25-35** | **>11** | 120 | **20-22** | **<1.8** | **2.2×1.0** |

## E. Circuit Layout Optimization Result

Obtained netlist of the circuit is then fed into optimization model. Firstly, the circuit is optimized using circuit schematic with initial parameters obtained from CoT. Then, the model replaces time-consuming EM simulation with schematic simulation and corrects the differences between them using ISM. During the entire optimization process, there are only 6 EM simulations for the layout and thousands of simulations for the schematic. As most of iterations are evaluated on schematic, the time is significantly reduced compared to the optimization using EM simulations. The results shown in Table III, the optimized LNA is superior than reported ones proving its efficiency.

## V. CONCLUSION

In this paper, we introduced DocEDA, an automated system that streamlines the extraction of electrical parameters from circuit datasheets and design documents. By leveraging advanced computer vision techniques and the reasoning capabilities of Large Language Models (LLMs), DocEDA addresses the inefficiencies and errors inherent in traditional manual extraction methods within Electronic Design Automation (EDA). The implementation of DocEDA demonstrates significant improvements in both the efficiency and accuracy of processing circuit design documents. Experimental results confirm that the system effectively reduces reliance on manual intervention, minimizes human error, and adapts to a variety of circuit design scenarios and document formats. This adaptability indicates DocEDA's potential to significantly impact and transform traditional methodologies in EDA.

ACKNOWLEDGMENT

This work is the integration of all authors' contribution. Hong Cai Chen designed the whole framework. Longchang Wu developed document parsing model, Yipin Xu proposed chain of thoughts, Jiarui Zhong constructed database, Ming Gao developed the image-to-circuit model, and Lingrui Shen proposed the circuit optimization algorithm.

REFERENCES

[1] H. Y. Wu, Z. L. He, X. Y. Zhang, *et al.*, "ChatEDA: A Large Language Model Powered Autonomous Agent for EDA," *IEEE Transactions on Computer-Aided Design of Integrated Circuits and Systems,* vol. 43, no. 10, pp. 3184-3197, Oct 2024.

[2] Y. Lai, S. Lee, G. Chen, *et al.*, "AnalogCoder: Analog Circuit Design via Training-Free Code Generation," *arXiv*: 2405.14918, 2024.

[3] C. Liu, W. Chen, A. Peng, *et al.*, "AmpAgent: An LLM-based Multi-Agent System for Multi-stage Amplifier Schematic Design from Literature for Process and Performance Porting," *arXiv*: 2409.14739, 2024.

[4] Y. Huang, T. Lv, L. Cui, *et al.*, " LayoutLMv3: Pre-training for Document AI with Unified Text and Image Masking," *30th ACM International Conference on Multimedia*, pp. 4083-4091, 2022.

[5] J. Li, Y. Xu, T. Lv, *et al.*, " DiT: Self-supervised Pre-training for Document Image Transformer," *A 30th ACM International Conference on Multimedia,* pp. 3530-3539, 2022.

[6] H. C. Chen, L. Wu, and Y. Zhang, "EDocNet: Efficient Datasheet Layout Analysis Based on Focal and Global Knowledge Distillation," *Under review*.

[7] H. C. Chen, Y. P. Xu, and Y. Zhang, "Advanced Chain-of-Thought Reasoning for Parameter Extraction from Documents Using Large Language Models," *Under review*.

[8] J. Wang, M. Pan, T. He, *et al.*, "A pseudo-relevance feedback framework combining relevance matching and semantic matching for information retrieval," *Information Processing & Management*, vol. 57, no. 6, p. 102342, Nov. 2020.

[9] Y. Qi, J. Zhang, W. Xu, *et al.*, "Salient context-based semantic matching for information retrieval," *EURASIP Journal on Advances in Signal Processing*, vol. 2020, no. 1, pp. 1–17, Dec. 2020.

[10] M. Gao, R. Qiu, Z. H. Chang, K. Zhang, H. Wei, and H. C. Chen, "Circuit Diagram Retrieval Based on Hierarchical Circuit Graph Representation,". *Under review*.

[11] R. Varghese and M. S, "YOLOv8: A Novel Object Detection Algorithm with Enhanced Performance and Robustness," in 2024 International Conference on Advances in Data Engineering and Intelligent Computing Systems (ADICS), pp. 1-6, 2024.

[12] Y. Liu, Z. Shao, and N. Hoffmann, "Global Attention Mechanism: Retain Information to Enhance Channel-Spatial Interactions," *ArXiv*: 2112.05561, 2021.

[13] J. Zhang, X. Yan, H. Luo and Y. Guo, "An Efficient Transistor-Model-Assisted Layout Synthesis Approach Using Improved Implicit Space Mapping for High-Performance MMIC PAs," *IEEE Transactions on Circuits and Systems II: Express Briefs*, vol. 71, no. 5, pp. 2639-2643, May 2024.

[14] A. Zhao, T. Gu, Z. Bi, *et al.*, "D3PBO: Dynamic Domain Decomposition-based Parallel Bayesian Optimization for Large-scale Analog Circuit Sizing". *ACM Trans. Des. Autom. Electron. Syst.* vol. 29, no, 3, pp. 1-25, 2024.

[15] D. Zhan. "Expected Coordinate Improvement for High-Dimensional Bayesian Optimization," *Swarm and Evolutionary Computation.* 2024, 91, 101745.

[16] L. Yang, L. -A. Yang, T. Rong, Y. Li, Z. Jin and Y. Hao, "Codesign of Ka-Band Integrated GaAs PIN Diodes Limiter and Low Noise Amplifier," *IEEE Access*, vol. 7, pp. 88275-88281, 2019.